\documentclass[final,twocolumn,3p,times]{elsarticle}

\usepackage{CJK}%
\usepackage{color}%
\usepackage{amsmath,bm}%
\usepackage{graphicx}%
\usepackage{dcolumn}
\usepackage{longtable}
\usepackage{array}
\usepackage{lineno}

\journal{Journal of Physics G}

\begin{document}

\begin{frontmatter}

\title{Investigation of the nuclear liquid-gas phase transition in the static AMD}

\author[label1]{W. Lin}
\author[label1]{P. Ren}
\author[label1]{X. Liu\corref{cor1}}
\cortext[cor1]{Corresponding author.
	Email address: liuxingquan@scu.edu.cn (X. Liu)}
\author[label2]{H. Zheng}
\author[label3]{M. Huang}
\author[label1]{G. Qu}
\author[label4,label5]{R. Wada}
\address[label1]{Key Laboratory of Radiation Physics and Technology of the Ministry of Education, Institute of Nuclear Science and Technology, Sichuan University, Chengdu 610064, China}
\address[label2]{School of Physics and Information Technology, Shaanxi Normal University, Xi'an 710119, China}
\address[label3]{College of Physics and Electronics information, Inner Mongolia University for Nationalities, Tongliao, 028000, China}
\address[label4]{Cyclotron Institute, Texas A$\&$M University, College Station, Texas 77843}
\address[label5]{School of Physics, Henan Normal University, Xinxiang 453007, China}

\begin{abstract}
Nuclear liquid-gas phase transitions are investigated in the framework of static antisymmetrized molecular dynamics (static AMD) model under either a constant volume or a constant pressure. A deuteron quadrupole momentum fluctuation thermometer is applied to extract the temperature of fragmenting systems of $^{36}$Ar and $^{100}$Sn. A plateau structure of caloric curves is observed under a constant volume for those system with a density $\rho \leq$ 0.03 fm$^{-3}$. A clear backbending in the caloric curves, which indicates a first order phase transition, is observed under a constant pressure with all pressures studied. The similar behavior of caloric curves of $^{36}$Ar and $^{100}$Sn systems indicates that there is no strong system size effect under a constant volume or a constant pressure. Both the mass distributions and the light particle multiplicities show a strong $\alpha$ clusterization at low excitation energies in the static AMD simulations. The liquid-gas phase transition measures of the multiplicity derivative (dM/dT) and the normalized variance of $Z_{max}$ (NVZ) are applied. The experimental caloric curves are also compared with those of $^{100}$Sn of the static AMD simulations under both the constant volume and the constant pressure conditions. Discussions are presented with the available experimental results and those from the static AMD simulations. Large errors in the experimental temperature measurements and those in the reconstruction technique for the primary fragmenting source hinder to draw a conclusion whether the phase transition occurs under either a constant volume or a constant pressure. This study suggests that different measures for the liquid-gas phase transitions should be examined besides the caloric curves in order to draw a conclusion.
\end{abstract}

\end{frontmatter}


\section*{I. Introduction}

Nuclear multifragmentation process was first predicted in 1930s~\cite{Bohr1936Nature} and has been extensively studied following the advent of 4$\pi$ detectors~\cite{Borderie2008PPNP,Gulminelli2006EPJA,Chomaz2004PR,Scharenberg2001PRC,Xi1997ZPA}. It provides a wealth of information on nuclear dynamics, properties of the nuclear equation of state (EOS), possible nuclear liquid-gas phase transition among others. The nuclear liquid-gas phase transition was first studied in the early 1980s~\cite{Finn1982PRL,Minich1982PLB,Hirsch1984NPA} and has long been a hot topic of contemporary nuclear physics. It was suggested in the experimental observations and theoretical simulations, due to the resemblance between the equation of state of homogeneous nuclear matter and homogeneous Van der Waals matter. 

In the past four decades, many experimental and theoretical works have been devoted to searching for the signals of the liquid-gas phase transition in the Fermi energy heavy-ion collisions and relativistic energy projectile fragmentations. Measures used for the studies are the nuclear specific heat capacity (the caloric curves)~\cite{Suraud1989PPNP,Bonche1984NPA,Gross1993PPNP,Hagel1988NPA,Wada1989PRC,Cussol1993NPA,Pochodzalla1995PRL,Wada1997PRC,Hagel2000PRC,Natowitz2002PRC,Furuta2006PRC,Furuta2009PRC,Gupta2001ANP,ChomazPRL2000,Borderie2013PLB}, the bimodality in charge  asymmetry~\cite{Lopez2005PRL,Pichon2006NPA,Fevre2008PRL,Fevre2009PRC,Bonnet2009PRL,Borderie2010NPA}, the Fisher droplet model analysis~\cite{Fisher1967RPP,Elliott2002PRL,Ma2005PRC,Ma2005NPA,Ma2004PRC,Huang2010PRC,Giuliani2014PPNP,Lin2014PRC}, the Landau free energy approach~\cite{Bonasera2008PRL,Huang2010PRC,Giuliani2014PPNP,Tripathi2011PRC,Tripathi2011JPCS,Tripathi2012IJMPE,Mabiala2013PRC}, the moment of the charge distributions~\cite{Campi1988PLB,Campi1986JPA,Das1998PRC,Mastinu1998PRC,Ma2005PRC}, the fluctuation properties of the heaviest fragment size (charge)~\cite{Mastinu1998PRC,Ma2005PRC,Ma2005NPA,Botet2001PRL,Frankland2005PRC,Souza2020Arxiv}, the Zipf's law~\cite{Ma1999PRL,Ma1999EPJA}, the Shannon information entropy~\cite{Ma1999PRL,CWMa2018PPNP}, the spinodal decomposition analysis~\cite{Borderie2001PRL,Borderie2018PLB}, the multiplicity derivatives proposed for a signature of first order phase transition~\cite{Mallik2017PRCR,Gupta2018PRC} and the derivative of cluster size~\cite{Das2018PLB}. The liquid drop model parameters~\cite{Chaudhuri2019PRC}, the clusterization algorithms~\cite{Sood2019PRC} effects on the liquid-gas phase transition, the finite-size scaling phenomenon~\cite{Liu2019PRC}, the critical parameters and the microscopic predictions of the liquid-gas phase transition were also studied recently~\cite{Yang2019PRC,Carbone2018PRC}. Many considerable progresses have been accomplished on the theoretical and experimental studies.

In our recent studies~\cite{Lin2018PRC2,Lin2019PRC}, several measures among listed above were investigated and the solidarity of the signal of these measures for different sizes, N/Z asymmetries and volumes of the fragmenting system were examined in the framework of the statistical multifragmentation model (SMM)~\cite{ZhangNPA1987I,ZhangNPA1987II,Bondorf95,Botvina2001,Soulioutis2007,Lin2018PRC}, in which the nuclear liquid-gas phase transition occurs in a constant volume. In the references, a clear signal for the first order phase transition is observed for the examined measures, and the charge distributions for different system sizes provide an instructive picture for the signals observed in the SMM events. No system size effect is observed in them. This is quite different from the results by the percolation model or the lattice gas model, in which the phase transition signatures become prominent when the system size increases~\cite{Elliott1994PRC,Li1994PRC,Moretto2005PRL,Elliott1997PRC}. These measures were also experimentally studied using reactions of $^{40}$Ar+$^{27}$Al, $^{48}$Ti and $^{58}$Ni at 40 A MeV and clear signals for a possible liquid-gas phase transition were observed~\cite{Wada2019PRC}. 

In the SMM adopted in our previous studies of Refs.~\cite{Lin2018PRC2,Lin2019PRC}, a constant volume is assumed for the breakup system and a spherical volume with the normal density is assigned for the fragments, and their kinetic energies are determined from the energy balance. In this article, these studies are extended in the framework of static antisymmetrized molecular dynamics (static AMD) model~\cite{Furuta2006PRC,Furuta2009PRC}, where the multi-fragmentation is treated in a quantum branching process and the fragments can be deformed in the phase space.
This article is organized as follows: Brief descriptions of the static AMD and the deuteron quadrupole momentum fluctuation thermometer are presented in Sec. II. The caloric curves under constant volume and constant pressure conditions are presented in Sec. III. The liquid-gas phase transition measures in the static AMD are shown in Sec. IV. Comparisons with the available experimental data are carried out in Sec. V. A brief summary is given in Sec. VI.

\section*{II. Static AMD model and deuteron quadrupole momentum fluctuation thermometer}

In AMD~\cite{Ono1992PRL,Ono1992PTP,Ono1996PRC,Ono2002PRC,Ono2004PPNP,Ono2004PRCR}, the reaction system with N nucleons is described as a Slater determinant of N Gaussian wave packets,
\begin{eqnarray}\label{Eq1}
\Phi(Z)=\det\left[\exp\left\{-\nu\left(\textit{\textbf{r}}_{j}-\frac{\textit{\textbf{Z}}_{i}}{\sqrt{\nu}}\right)^{2}+\frac{1}{2}\textit{\textbf{Z}}_{i}^{2}\right\}\chi_{\alpha_{i}}(j)\right],
\end{eqnarray}
where the complex variables $Z\equiv\{\textit{\textbf{Z}}_{i};i=1,\ldots,N\}=\{Z_{i\sigma};i=1,\ldots,N, \sigma=x,y,z\}$ represent the centroids of the wave packets. $\chi_{\alpha_{i}}$ represents the spin and isospin states of p$\uparrow$, p$\downarrow$, n$\uparrow$, or n$\downarrow$. The width parameter $\nu$ is taken as $\nu =$ 0.16 fm$^{-2}$ in order to reproduce the binding energy of nuclei. The experimental binding energies are reproduced within 10\% for most nuclei~\cite{Ono2004PRCR}. Using the centroid of the Gaussian wave packets, the time evolution of $Z$ is determined classically by the time-dependent variational principle and the two body nucleon collision process. The equation of motion is described as,
\begin{eqnarray}\label{Eq2}
i\hbar\sum_{j\tau}C_{i\sigma,j\tau}\frac{dZ_{j\tau}}{dt}=\frac{\partial  \mathcal{H}}{\partial Z^{*}_{i\sigma}}.
\end{eqnarray}
Here $C_{i\sigma,j\tau}$ is a Hermitian matrix defined by $C_{i\sigma,j\tau}=\frac{\partial^{2}}{\partial Z^{*}_{i\sigma} \partial Z_{j\tau}}\log\left\langle\Phi(\textbf{Z})|\Phi(\textbf{Z})\right\rangle$. $\mathcal{H}$ is the expectation value of the Hamiltonian after the subtraction of the spurious kinetic energy of the zero-point oscillation of the fragment center of masses,
\begin{eqnarray}\label{Eq3}
\mathcal{H}=\frac{\left\langle\Phi(\textbf{Z})|H|\Phi(\textbf{Z})\right\rangle}{\left\langle\Phi(\textbf{Z})|\Phi(\textbf{Z})\right\rangle}-\frac{3\hbar^{2}\nu}{2M}A+T_{0}[A-N_{F}(Z)],
\end{eqnarray}
where $M$ is the mass of nucleons, $A$ is the mass number of system, $N_F(Z)$ is the fragment number, and $T_0$ is $3\hbar^2\nu/2M$ in principle but treated as a free parameter for the adjustment of the binding energies.
The wave packet diffusion process~\cite{Ono1996PRC} is taken into account stochastically in the time evolution of the wave packets in order to make a proper multifragmentation of hot nuclear matter generated during collisions.

Similar to other transport models, there are two separated processes, one is the mean field propagation of nucleons and the other is the nucleon-nucleon (NN) collision process. The mean field propagation is governed by a given effective interaction and the NN collision rate is determined by a given NN cross section. Pauli principle is fully respected in an exact manner in both processes. Throughout this paper, the Gogny interaction~\cite{Gogny1980PRC} is used for the mean field.


In the static AMD~\cite{Furuta2006PRC,Furuta2009PRC}, the effective Hamiltonian $\mathcal{H}$ is modified for a better treatment of the gaseous nucleons, which is important for the temperature measurement. The $N_F(Z)$ is modified to take into account the number of isolated fragments and nucleons together. The potential energies between the gaseous nucleons are also modified by adding an additional term of $\mathcal{V}_{mod}(Z)$ in Eq. (\ref{Eq3}). The diffusion component of the original width of wave packet is also improved for the stochastic single particle motion to fully satisfy the energy conservation and to precisely evaluate the temperature.

Instead of using the NN collisions as the decoherence process, a different decoherence process is adopted in the static AMD, which takes place with the probability of 1/$\tau_0$ per unit time, where $\tau_0$ is a characteristic coherence time. This decoherence process affects all the nucleons located within the relative radius of 2 fm if more than three nucleons are found within this radius. Three different coherence time $\tau_0$ (250, 500 and 1000 fm/c) were studied in Ref.~\cite{Furuta2006PRC} and found that they correspond to different level density parameter $a$ for the liquid caloric curve. Later the decoherence process was further improved and $\tau_0$ was readjusted as 5 fm/c in Ref.~\cite{Furuta2009PRC}. Therefore, the same parameter $\tau_0 = $ 5 fm/c is used throughout this paper.

An irregular surface reflection process at the container wall is introduced in the static AMD~\cite{Furuta2006PRC,Furuta2009PRC} to keep nucleons inside the container with a given radius $r_{wall}$. For each time step, nucleons i and j belong to the same cluster if the relative distance at the physical coordinate $|W_i - W_j| < 0.8$.
If an isolated nucleon $k$ is located outside the container and its momentum $\textit{\textbf{p}}_k = 2\hbar\sqrt{\nu}Im\textit{\textbf{Z}}_k$ directs outward, an inward momentum direction $\textit{\textbf{\^{p}}}'_k = \textit{\textbf{p}}'_k/|\textit{\textbf{p}}'_k|$ is randomly chosen, which satisfies $\textit{\textbf{r}}_k \cdot \textit{\textbf{\^{p}}}'_k < 0$, where $\textit{\textbf{r}}_k = \frac{1}{\sqrt{\nu}}Re\textit{\textbf{Z}}_{k}$. The absolute value of the momentum $|\textit{\textbf{p}}'_k|$ is adjusted so as to conserve the total energy which is sometimes affected by antisymmetrization. The total angular momentum of the system is not conserved, which allows one to construct a microcanonical ensemble in a single run. A similar reflection procedure is applied to the center of mass coordinate of a cluster if any of the nucleon in the cluster is located outside the container wall to avoid the unphysical breaking of the cluster.

The microcanonical temperature is obtained from the average kinetic energy of nucleons in a gaseous subsystem in the static AMD as
\begin{eqnarray}\label{Eq4}
T=\frac{2}{3}\left\langle \frac{\mathcal{K}_G}{A_G} \right\rangle_{\left\lbrace A_G>0\right\rbrace } ,
\end{eqnarray}
where $\mathcal{K}_G$ and $A_G$ are the kinetic energy and the number of nucleons of gaseous subsystem, respectively. The gaseous subsystem is identified as nucleons for which the density of the nucleons with the same spin-isospin is sufficiently low ($\rho_G \leq \rho_0/200$, where $\rho_0 = 0.16$ fm$^{-3}$ is the saturation density of nuclear matter) and do not have more than one other nucleon within the relative distance of $r_G=3$ fm. The pressure is defined as the external force necessary to keep the volume and is given by
\begin{eqnarray}\label{Eq5}
	P=\frac{2}{4\pi r^2_{wall}\tau_{total}}\sum_{reflections}{\Delta \textit{\textbf{p}}\cdot\textbf{\^{n}}},
\end{eqnarray}
where the summation is taken over all the reflection nucleons and cluster at the container wall during the total evolution time $\tau_{total}$. $\Delta \textit{\textbf{p}}$ is the momentum change at each reflection and $\textbf{\^{n}}$ is the normal vector. The factor 2 comes from the fact that when a nucleon or fragment hits the wall, the total momentum of the rest system is also changed to conserve the momentum.

The static AMD calculations are performed with the excitation energy $E_x/A$ ranging from 2 MeV to 25 MeV with step of 0.5 MeV for $^{36}$Ar and $^{100}$Sn. The $r_{wall}$ ranging from 5 fm to 14.5 fm ($\rho$ from 0.00282 to 0.0688 fm$^{-3}$) is used for $^{36}$Ar and ranging from 7 fm to 20 fm ($\rho$ from 0.00296 to 0.0696 fm$^{-3}$) for $^{100}$Sn. For each $r_{wall}$ and excitation energy, the time evolution of AMD is calculated up to 55000 fm/c for $^{36}$Ar and 25000 fm/c for $^{100}$Sn. The output data are collected for the time step of 10 fm/c. Three individual runs are performed to increase the statistics.

\begin{figure}[hbt]
	\centering
	\includegraphics[scale=0.35]{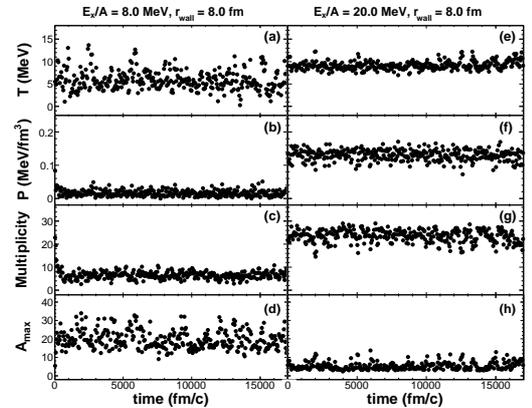}
	\caption{\footnotesize
	 	Time evolution of temperature (a) and (e), pressure (b) and (f), multiplicity (c) and (g), and maximum cluster mass (d) and (h) of $^{36}$Ar system. The left column corresponds to the system with $E_x/A=$ 8 MeV and $r_{wall}=$ 8 fm ($\rho =$ 0.0168 fm$^{-3}$). The right column is the results of the system with $E_x/A=$ 20 MeV and $r_{wall}=$ 8 fm. The average value of 5 output time steps is shown for each data point.
	}		
	\label{fig:fig01_time_evolution}
\end{figure}

Figure~\ref{fig:fig01_time_evolution} shows a typical example of the time evolution of temperature, pressure, multiplicity and maximum cluster mass in (a) and (e), (b) and (f), (c) and (g), and (d) and (h), respectively. The left and right columns correspond to the systems with the same $r_{wall}=$ 8 fm, but $E_x/A=$ 8 and 20 MeV, respectively. One can see that for the cases in Fig.~\ref{fig:fig01_time_evolution}, a short time is needed for systems to achieve a thermal equilibrium. The maximum time to achieve the thermal equilibrium is less than 4000 fm/c among the calculations mentioned above. Therefore, the same as that in Refs.~\cite{Furuta2006PRC,Furuta2009PRC}, the first 5000 fm/c states are discarded in the following analysis.

Several experimental thermometers such as the kinetic energy slope thermometer~\cite{Westfall1982PLB,Jacak1983PRL}, the isotopic ratio thermometer~\cite{Albergo1985NCA,Tsang1997PRL}, the population of excited state thermometer~\cite{Morrissey1984PLB,Pochodzalla1985PRL}, the fluctuation thermometer~\cite{Wuenschel2010NPA,Zheng2013PRC} and the temperature extracted by the Thomas-Fermi approach~\cite{Su2020JPG}, have been proposed in the past to evaluate the temperature of the hot nuclear matter. In the static AMD, the temperature of system is obtained from the average kinetic energy of nucleons in a gaseous subsystem. However, it is difficult in defining such a gaseous subsystem without the measured source velocity in nuclear reaction experiments.

In our previous work, a deuteron quadrupole momentum fluctuation thermometer was proposed in the experimental data analysis~\cite{Wada2019PRC}.
Using a classical Maxwell-Boltzmann distribution of momentum yields, a temperature from the quadrupole momentum fluctuation of a prob particle is derived by Wuenschel {\it et al.}~\cite{Wuenschel2010NPA} as
\begin{eqnarray}\label{Eq10}
T = \sqrt{\langle\sigma^2_{xy}\rangle/4m^2},
\end{eqnarray}
where $m$ is the probe particle mass, $\langle\sigma^2_{xy}\rangle$ is the variance of the two dimensional quadrupole momentum, which is defined as
\begin{eqnarray}\label{Eq11}
\nonumber Q_{xy} &=& p^2_x - p^2_y,\\
\langle\sigma^2_{xy}\rangle &=& \langle Q_{xy}^2\rangle - \langle Q_{xy}\rangle^2,
\end{eqnarray}
where $p_x$ and $p_y$ are the transverse components of the fragment momentum. According to the SMM simulation~\cite{Wada2019PRC}, the deuteron quadrupole momentum fluctuation thermometer is adopted to minimize the Coulomb and the secondary decay effects.

\begin{figure}[hbt]
	\centering
	\includegraphics[scale=0.35]{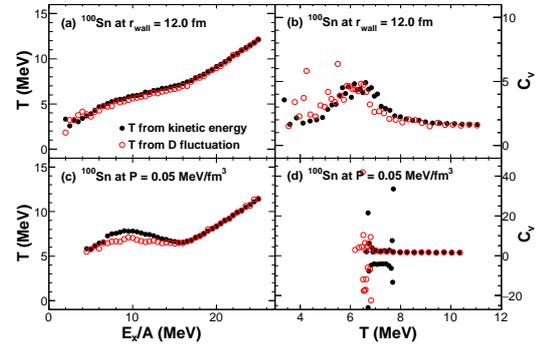}
	\caption{\footnotesize
		(Color online) (a) Caloric curves under a constant volume at the density $\rho=$ 0.0138 fm$^{-3}$ and (c) those under a constant pressure at $P$ = 0.05 MeV/fm$^{3}$ from static AMD with $^{100}$Sn. The derived specific heat capacity ($C_v$) are shown in (b) and (d) with the smoothing of seven points in the caloric curves. Open circles are for the deuteron quadrupole momentum fluctuation thermometer and solid circles are for the kinetic energy thermometer.
	}		
	\label{fig:fig02_Tk_TD}
\end{figure}

In Fig.~\ref{fig:fig02_Tk_TD}, the temperatures extracted by the deuteron quadrupole momentum fluctuation (open circles) and the average kinetic energy~\cite{Furuta2006PRC,Furuta2009PRC} (solid circles) are compared and good agreements are observed. To get the caloric curve under a constant pressure, the temperature is obtained from the interpolation of T vs P curve at a given pressure for each excitation energy. The caloric curves under a constant volume and under a constant pressure are shown in Fig.~\ref{fig:fig02_Tk_TD} (a) and (c), respectively, which will be discussed further in detail in the next section. The derived specific heat capacity ($C_v$) are shown in Fig.~\ref{fig:fig02_Tk_TD} (b) and (d) with smoothing over seven points in the caloric curves. Under a constant volume case, temperature increases rather smoothly as the excitation energy increases and one can only see a possible liquid-gas phase transition as a shoulder structure at $E_x/A =$ 8 - 15 MeV, which corresponds to the peak at $C_v$ around T = 6 MeV. On the contrary, under a constant pressure case, the liquid-gas phase transition is clearly observed as a backbending at $E_x/A =$ 8 - 15 MeV, which corresponds to the negative $C_v$ between T = 6 - 8 MeV. Slightly larger fluctuation is observed for the deuteron quadrupole momentum fluctuation thermometer values in the lower excitation energy side, because deuterons have less probability at these low excitation energies. Since the good agreement of the temperature obtained by the deuteron quadrupole momentum fluctuation and the average kinetic energy, the deuteron quadrupole momentum fluctuation thermometer is used in the following analysis. It is adopted also for the direct comparisons between the AMD simulations and experimental observations in future.

\section*{III. Caloric curves under constant volume and constant pressure}

Sobotka once argued whether one should use the Helmholtz free energy (under a constant volume) or the Gibbs free energy (under a constant pressure) to characterize the hot nuclear matter observables~\cite{Sobotka2011PRC}. Experimentally, when a nuclear liquid-gas phase transition occurs during the expansion, whether it takes place under a constant volume or under a constant pressure is difficult to distinguish. Here we propose a possible scenario to distinguish these processes, using the static AMD simulated events of $^{36}$Ar and $^{100}$Sn.
 
\begin{figure}[htb]
	\centering
	\includegraphics[scale=0.35]{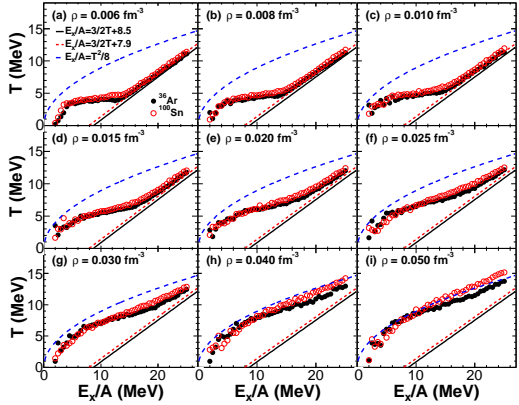}
	\caption{\footnotesize
		(Color online) Caloric curves of $^{36}$Ar (solid circles) and $^{100}$Sn (open circles) under the constant volume of density $\rho$ = 0.006 to 0.050 fm$^{-3}$ from (a) to (i). The solid, dashed and long dashed lines correspond to $E_{x}/A = 3/2T-E_{g.s.}/A$ of $^{36}$Ar and $^{100}$Sn and the Fermi gas formula $E_{x}/A = aT^{2}$ with $a$ = 1/8, respectively. The large fluctuation of data at lower excitation energies is caused by the low statistics.
	}		
	\label{fig:fig03_Caloric_volume}
\end{figure}

The calculations are done in discrete $r_{wall}$ for both $^{36}$Ar and $^{100}$Sn systems. It is difficult to find exactly the same density for $^{36}$Ar and $^{100}$Sn at the $r_{wall}$ selected. Therefore, similar to the constant pressure case, the temperature for a given density is obtained from the interpolation of T vs $\rho$ curve for each excitation energy. Figure~\ref{fig:fig03_Caloric_volume} shows the caloric curves of $^{36}$Ar (solid circles) and $^{100}$Sn (open circles) under the constant volume of nine different densities. The lines correspond to $E_{x}/A = 3/2T-E_{g.s.}/A$, ($E_{g.s.}/A$ = -8.5 MeV for $^{36}$Ar and -7.9 MeV for $^{100}$Sn), and the dashed curves correspond to $E_{x}/A = aT^{2}$ with $a$ = 1/8 are shown for comparison. One can see from Fig.~\ref{fig:fig03_Caloric_volume} that the trend of the caloric curves for $^{36}$Ar and $^{100}$Sn are very similar for a given density. A degraded plateau structure is found for densities $\rho >$ 0.03 fm$^{-3}$, which indicates the liquid-gas phase transition happens below the densities $\rho =$ 0.03 fm$^{-3}$. Above this density, both the two systems stay in liquid phase all the time and the caloric curves follow well the curve of $E_x/A = aT^2$. A plateau structure is developed as the density decreases. As discussed in our previous work~\cite{Lin2019PRC}, the similar plateau temperatures at a given density ($\rho \leq$ 0.03 fm$^{-3}$) between the two systems reflect a negligible dependence of the phase transition temperature on the system size.

\begin{figure}[hbt]
	\centering
	\includegraphics[scale=0.35]{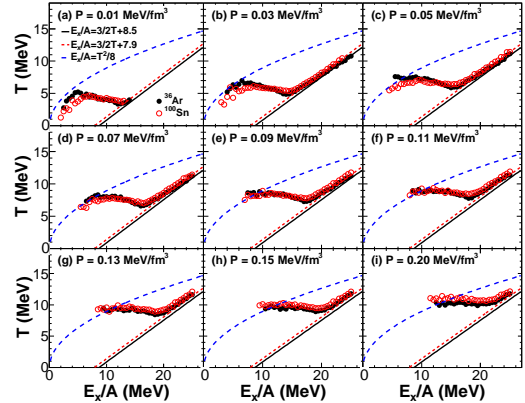}
	\caption{\footnotesize
		(Color online) Caloric curves of $^{36}$Ar (solid circles) and $^{100}$Sn (open circles) under the constant pressure of $P$ = 0.01 to 0.20 MeV/fm$^{3}$ from (a) to (i). The solid, dashed and long dashed lines are same as those in Fig.~\ref{fig:fig03_Caloric_volume}. The large fluctuation of data at lower excitation energies is caused by the low statistics.
	}		
	\label{fig:fig04_Caloric_pressure}
\end{figure}

Similar plots for the caloric curves of $^{36}$Ar (solid circles) and $^{100}$Sn (open circles) at constant pressure of $P = 0.01$ to 0.2 MeV/fm$^{3}$ are shown in Fig.~\ref{fig:fig04_Caloric_pressure} (a) to Fig.~\ref{fig:fig04_Caloric_pressure} (i). As shown in Fig.~\ref{fig:fig04_Caloric_pressure} a clear backbending is observed in the caloric curves at each given pressure. This is a distinct difference from those under a constant volume. The backbendings are observed similarly under all pressures examined. At lower excitation energies, the systems are in the liquid phase and no gaseous nucleons or light fragments are produced, causing the left end of the calculated caloric curves stop at the dashed curve. For a given pressure, the phase transition temperatures are similar for $^{36}$Ar and $^{100}$Sn, which again indicates no system size effect for the phase transition temperature under the constant pressure condition in the static AMD simulations.

\section*{IV. Liquid-gas phase transition measures in static AMD}

Several experimental measures were examined in our previous works in the framework of SMM model~\cite{Lin2018PRC2,Lin2019PRC}. The multiplicity derivative (dM/dT) and the normalized variance of $Z_{max}$ (NVZ) were found to provide accurate measures for the liquid-gas phase transition temperature from both the primary and the secondary fragments. In this section, the mass distributions, the multiplicities of light particles, the dM/dT and the NVZ are examined for the static AMD calculations.

\begin{figure}[hbt]
	\centering
	\includegraphics[scale=0.35]{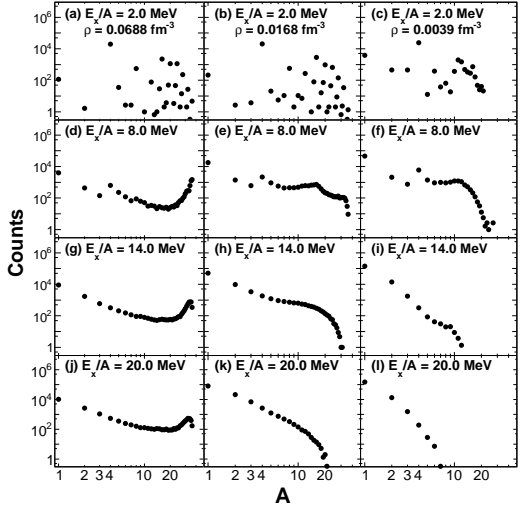}
	\caption{\footnotesize
		The mass distributions of $^{36}$Ar system at $\rho=$ 0.0688 fm$^{-3}$ ($r_{wall} =$ 5 fm) (left column), 0.0168 fm$^{-3}$ ($r_{wall} =$ 8 fm) (middle column) and 0.0039 fm$^{-3}$ ($r_{wall} =$ 13 fm) (right column) for the excitation energies $E_x/A =$ 2, 8, 14 and 20 MeV from top to bottom.
	}		
	\label{fig:fig05_mass_distribution_Ar36}
\end{figure}

The mass distributions of $^{36}$Ar system are shown in Fig.~\ref{fig:fig05_mass_distribution_Ar36} for $\rho=$ 0.0688 fm$^{-3}$ ($r_{wall} =$ 5 fm, left column), 0.0168 fm$^{-3}$ ($r_{wall} =$ 8 fm, middle column) and 0.0039 fm$^{-3}$ ($r_{wall} =$ 13 fm, right column) and for the excitation energies $E_x/A =$ 2, 8, 14 and 20 MeV from top to bottom. The fragments are identified by the relative distance of nucleons $r_{clust} \leq$ 2.5 fm in the coordinate space. At low excitation energy of $E_x/A=$ 2 MeV, the mass distributions show large fluctuations for different mass number A and the maximum yield appears at A = 4 for all the densities investigated. When system evolves at a higher density, the U shape distribution is found even at the excitation energy $E_x/A=$ 20 MeV. For the $\rho=$ 0.0168 fm$^{-3}$, the system evolves from two Ar like peaks to the power law distribution as the $E_x/A$ increases. The system at $\rho =$ 0.0039 fm$^{-3}$ evolves more rapidly as the $E_x/A$ increases. As pointed out in Ref.~\cite{Furuta2006PRC}, the mass distribution depends significantly on the choice of the decoherence parameter $\tau_{0}$ and the cluster radius $r_{clust}$. Nevertheless, it is helpful for the qualitative understanding of the evolution of the system under a constant volume and a constant pressure, together with the results in Ref.~\cite{Furuta2006PRC}.

\begin{figure}[hbt]
	\centering
	\includegraphics[scale=0.35]{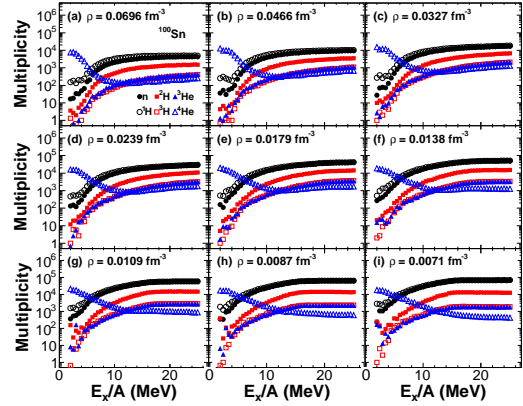}
	\caption{\footnotesize
		(Color online) Light particle multiplicities of $^{100}$Sn system as a function of the excitation energy for $r_{wall}=$ 7 to 15 fm in (a) to (i), corresponding to density $\rho=$ 0.0071 to 0.0696 fm$^{-3}$. The solid circles, open circles, solid squares, open squares, solid triangles and open triangles represent n, $^{1}$H, $^{2}$H, $^{3}$H, $^{3}$He and $^{4}$He, respectively.
	}		
	\label{fig:fig06_LPs_multiplicity_Sn100}
\end{figure}

The $\alpha$ cluster enhancement at low density has been observed experimentally from NIMROD data~\cite{Qin2012PRL}. In the static AMD calculations, on the other hand, the enhancement of the $\alpha$ cluster yield is also found at lower excitation energies independent of the wall radius. It is interesting to investigate the evolution of light particle multiplicities as a function of the excitation energy. The light particle multiplicities of $^{100}$Sn system as a function of the excitation energy are shown in Fig.~\ref{fig:fig06_LPs_multiplicity_Sn100} for different densities. One can see from Fig.~\ref{fig:fig06_LPs_multiplicity_Sn100} that a strong $\alpha$ clusterization is found at liquid or liquid-gas coexist region (low excitation energy side) for all the densities investigated. The multiplicity of $\alpha$ cluster increases only slightly when the density decreases at low excitation energies, which is a significant contrast to the strong $\alpha$ cluster enhancement at low densities observed in the experiment~\cite{Qin2012PRL}. The multiplicities of the other light particles increase smoothly as the excitation energy increases. The $\alpha$ yield enhancement starts close to the temperature where the liquid-gas phase transition occurs when the temperature decreases, but the role of this $\alpha$ clusterization for the liquid-gas phase transition is not understood yet.

\begin{figure}[hbt]
	\centering
	\includegraphics[scale=0.35]{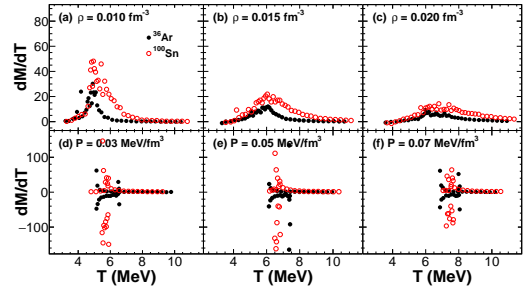}
	\caption{\footnotesize
		(Color online) dM/dT as a function of the temperature T for the constant density of $\rho =$ 0.01 fm$^{-3}$ (a), 0.015 fm$^{-3}$ (b) and 0.02 fm$^{-3}$ (c) as well as for the constant pressure of $P =$ 0.03 MeV/fm$^{3}$ (d), 0.05 MeV/fm$^{3}$ (e) and 0.07 MeV/fm$^{3}$ (f). Solid and open circles represent the system of $^{36}$Ar and $^{100}$Sn. The data are obtained with the smoothing of seven points of multiplicity and temperature.
	}		
	\label{fig:fig07_dMdT}
\end{figure}

The multiplicity derivatives (dM/dT) of $^{36}$Ar and $^{100}$Sn systems are shown in Fig.~\ref{fig:fig07_dMdT} under the constant volume of density $\rho =$ 0.01, 0.015 and 0.02 fm$^{-3}$ in (a) (b) and (c), respectively, as well as those under the constant pressure of $P =$ 0.03, 0.05 and 0.07 MeV/fm$^{3}$ in (d), (e) and (f), respectively. One should note that the resemblance between $C_v$ values in Figs.~\ref{fig:fig02_Tk_TD} (b), (d) and dM/dT values Figs.~\ref{fig:fig07_dMdT}(b), (e)  where the same density and pressure are used. The characteristic peak of dM/dT under a constant volume are found for all the three densities investigated. However, the peak temperature increases and the width of the peak becomes broader as the density increases under the constant volume condition, which is quite different from that of SMM calculations in our previous work~\cite{Lin2019PRC}, where the peak temperature stays same and the width becomes narrower as the density increases. The reason could be the assumption of a spherical shape of fragments used in SMM, whereas in the static AMD, no shape assumption is used, which affects in the formation of clusters in the excitation energy and size. Similar to the $C_v$ under the constant pressure condition in Fig.~\ref{fig:fig02_Tk_TD} (d), the negative dM/dT is observed for both $^{36}$Ar and $^{100}$Sn systems under the constant pressure condition. The region of negative dM/dT shifts to high temperature when the pressure increases. A narrower region of negative dM/dT is also found for $^{100}$Sn system, which reflects a sharper liquid-gas phase transition for larger system size under a constant pressure condition.

\begin{figure}[hbt]
	\centering
	\includegraphics[scale=0.35]{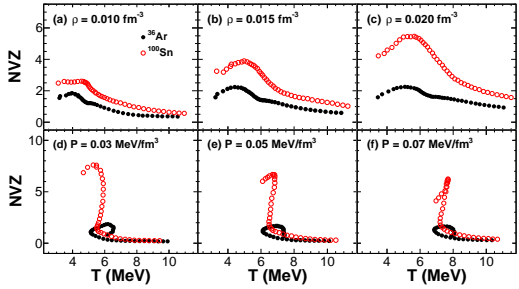}
	\caption{\footnotesize
		(Color online) The same as that in Fig.~\ref{fig:fig07_dMdT} but for the normalized variance of $Z_{max}$ (NVZ) as a function of temperature. The data are obtained with the smoothing of seven points of NVZ and temperature.
	}		
	\label{fig:fig08_NVZ}
\end{figure}

At the temperature of the liquid-gas phase transition, the system shows the maximum fluctuation~\cite{Ma2005PRC,Ma2005NPA,Ma2004PRC}. In order to see such a feature, the normalized variance of $Z_{max}$ (NVZ) are also investigated for both $^{36}$Ar and $^{100}$Sn systems. Figure~\ref{fig:fig08_NVZ} shows the NVZ as a function of temperature under constant volume of density $\rho =$ 0.01, 0.015 and 0.02 fm$^{-3}$ in (a) (b) and (c), respectively, as well as those under constant pressure of $P =$ 0.03, 0.05 and 0.07 MeV/fm$^{3}$ in (d), (e) and (f), respectively. The maximum value of NVZ appears at different temperatures, and these peaks appear at the temperatures lower than that from the dM/dT with significant broadening. Under the constant pressure, on the contrary, much sharper signals are observed for both systems for the liquid-gas phase transition. The maximum value of NVZ agrees well with each other for the two systems investigated under a constant pressure condition. The temperature increases and they correspond more or less to those of dM/dT in the lower panels of Fig.~\ref{fig:fig07_dMdT}, especially for $^{100}$Sn. A slight broadening of the signatures is observed for $^{36}$Ar system.

In these comparisons with different measures, all indicate that the temperatures for the liquid-gas phase transitions depend on the density and pressure, but no system size effect is observed. Under a constant pressure, the signal becomes much sharper at the liquid-gas phase transition temperature, which corresponds to the backbending of the caloric curves in Fig.~\ref{fig:fig04_Caloric_pressure}.

\section*{V. Comparison with the available experimental data}

In general, in nuclear collisions, the particle emitting source is expanding and therefore the volume and pressure is changing in time. However in  the studies of Furuta and Ono in Ref.~\cite{Furuta2009PRC}, they can identify the mass distribution at a given time from the dynamical simulation of AMD to one of the ensembles from the static AMD. This suggests that for a given system and a given excitation energy, the maximum fluctuation occurs at a certain volume or a certain pressure. As seen in the previous two sections, from the distinctive features of the measures under a constant volume or a constant pressure for the liquid-gas phase transition, we may be able to distinguish these two conditions using the results from the heavy ion experiments. In this section, we will compare our static results with those obtained from the heavy ion collisions.

\begin{figure}[htb]
	\centering
	\includegraphics[scale=0.35]{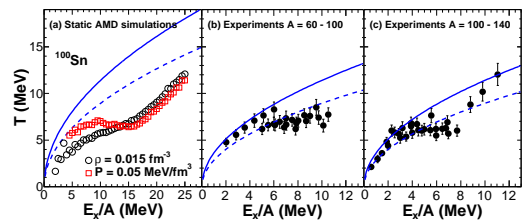}
	\caption{\footnotesize
		(Color online) (a) Caloric curves of $^{100}$Sn under a constant volume of $\rho$ = 0.015 fm$^{-3}$ (open circles) and a constant pressure of $P$ = 0.05 MeV/fm$^{3}$ (open squares). (b) Experimental caloric curves of source mass number A = 60 - 100 and (c) those for A = 100 - 140. The experimental data are taken from Ref.~\cite{Natowitz2002PRC}. The solid and dashed curves correspond to the Fermi gas formula $E_{x}/A = aT^{2}$ with $a$ = 1/13 and 1/8, respectively.
	}		
	\label{fig:fig09_Exp_StaticAMD_Sn100}
\end{figure}

The available experimental data of caloric curves extracted from the isotope ratio and the He slope thermometers of the source mass ranges of A = 60 - 100 and 100 - 140~\cite{Natowitz2002PRC} are first compared with the present results of $^{100}$Sn. Experimentally, the difference in the experimental filter, the event classification and the methods applied to extract the temperature and the excitation energy may present quite different caloric curves. It is not our goal to go deep into the detail of the extraction of caloric curve, but rather we focus on the comparison of the trend of the caloric curve from the experiments and that from the static AMD simulation, expecting to reveal some clue for the liquid-gas phase transition.

Figure~\ref{fig:fig09_Exp_StaticAMD_Sn100} (a) shows the comparison of caloric curves of $^{100}$Sn in the static AMD simulations under a constant volume of $\rho$ = 0.015 fm$^{-3}$ (open circles) and under a constant pressure of $P$ = 0.05 MeV/fm$^{-3}$ (open squares). The experimental caloric curves of source mass number A = 60 - 100 and 100 - 140 are shown in Fig.~\ref{fig:fig09_Exp_StaticAMD_Sn100} (b) and (c). The experimental data are taken from Ref.~\cite{Natowitz2002PRC}, in which the published data are sorted out in different source sizes. One should note that the scale of the x-axis is different between the simulation and the experimental data. The two caloric curves of $^{100}$Sn with phase transition temperature $T\sim$ 6 MeV from the AMD simulations are compared to the experimental values. In the experimental data, a plateau appears at $E_x/A \sim$ 3 to 5 MeV and last up to 10 MeV or more. The three points at highest excitation energy for A = 100 - 140 in the experimental data are not observed for A = 60 - 100 results, indicating more experimental data are necessary for $E_x/A >$ 10 MeV. No clear backbending is observed in the experimental results within the experimental accuracy. In the case of constant volume in the simulation as shown in Fig.~\ref{fig:fig03_Caloric_volume}, the plateau appears at $E_x/A =$ 5 - 10 MeV and last up to $\sim$ 15 MeV. On the other hand, in the case of the constant pressure, the plateau starts at excitation energy $E_x/A$ less than 5 MeV and a backbending appears between 10 - 15 MeV. But by simply looking at the results in Figs.~\ref{fig:fig03_Caloric_volume} and \ref{fig:fig09_Exp_StaticAMD_Sn100}, the experimental results seem to favor that the liquid-gas phase transition occurs under a constant or approximately constant pressure. Unfortunately, this conclusion is not conclusive within the current experimental accuracy.

\begin{figure}[htb]
	\centering
	\includegraphics[scale=0.35]{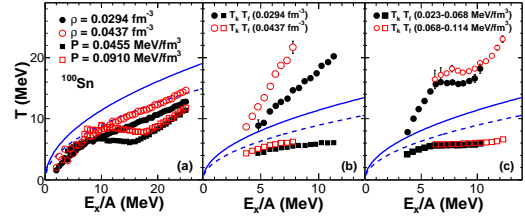}
	\caption{\footnotesize
		(Color online) (a) Caloric curves of $^{100}$Sn under a constant volume of $\rho$ = 0.0294 fm$^{-3}$ (solid circles) and 0.0437 fm$^{-3}$ (open circles) and under a constant pressure of $P$ = 0.0455 MeV/fm$^{3}$ (solid squares) and 0.091 MeV/fm$^{3}$ (open squares). (b) Experimental caloric curves under constant volume of $\rho$ = 0.0294 fm$^{-3}$ (solid symbols) and 0.0437 fm$^{-3}$ (open symbols). (c) Experimental caloric curves under constant pressure of $P$ = 0.0455 MeV/fm$^{3}$ (solid symbols) and 0.091 MeV/fm$^{3}$ (open symbols). Circles represent temperature obtained with the kinetic energy and squares are those temperature obtained with the internal fragment temperature. The experimental data are taken from Ref.~\cite{Borderie2013PLB}. The solid and dashed curves correspond to the Fermi gas formula $E_{x}/A = aT^{2}$ with $a$ = 1/13 and 1/8, respectively.
	}		
	\label{fig:fig10_Borderie_Exp_StaticAMD_Sn100}
\end{figure}

In Ref.~\cite{Borderie2013PLB}, Borderie et al. analyzed the $^{129}$Xe+$^{nat}$Sn reactions at 32 - 50 MeV/nucleon from the INDRA detector. In their analysis the primary fragments at the freeze-out volume were reconstructed in an event-by-event basis. The temperature with the kinetic energy thermometer ($T_{k}$) and the internal fragment thermometer ($T_{f}$), the density and the pressure of the reconstructed events were obtained. The obtained caloric curves under constant volumes for the freeze-out volume $V_{FO}=$ 3.66 and 5.44 $V_0$, which correspond to density $\rho=$ 0.0437 and 0.0294 fm$^{-3}$, and under constant pressures of $P_{FO}=$ 0.023-0.068 and 0.068-0.114 MeV/fm$^{3}$ are shown in Fig.~\ref{fig:fig10_Borderie_Exp_StaticAMD_Sn100} (b) and (c), respectively. The caloric curves of the static AMD calculations of $^{100}$Sn system are plotted in Fig.~\ref{fig:fig10_Borderie_Exp_StaticAMD_Sn100} (a) for comparisons. Even though the system size is around 200 for those in Fig.~\ref{fig:fig10_Borderie_Exp_StaticAMD_Sn100} (b) and (c), these comparisons are reasonable because no strong system size effect is observed in the static AMD results. The caloric curves with the internal fragment thermometer are similar to those of the static AMD, but show less dependence on the density or pressure. Only a flat plateau is observed in their result under a constant pressure. Their results with the kinetic energy thermometer seems problematic, since the temperatures become far beyond the Fermi-gas curves (blue solid and dashed curves), where a clear backbending is observed under the constant pressure. It is possible that the Coulomb or the dynamic effects were not fully removed in the reconstruction process. As suggested in this article, it is important to cross-check their results, using other measures with their freeze-out particles.

With the discussions above, the experimental uncertainties in the former and the ambiguity in the freeze-out event reconstruction in the latter certainly prevent to draw a conclusion. Therefore, more precise measurements of caloric curves in heavy ion collision experiments are still needed. Different measures other than those studied in the last section may provide additional evidence to the liquid-gas phase transition. Therefore, different measures for the liquid-gas phase transitions should be examined besides the caloric curves to draw a conclusion.

It should be noted that the plateau temperature of the experimental caloric curves decreases as the system size increases as shown in Ref.~\cite{Natowitz2002PRC}. The caloric curves extracted by the deuteron quadrupole momentum fluctuation from our recently analysis using reactions of $^{40}$Ar+$^{27}$Al, $^{48}$Ti and $^{58}$Ni at 40 A MeV~\cite{Wada2019PRC} also agree with that from previous experiments of the system size A = 30 - 60 using isotope ratio thermometer and He slope thermometer~\cite{Natowitz2002PRC}. This system size effect is not observed in the static AMD or the SMM simulations. It could be the effects of the dynamic process during the heavy ion collisions, which still need further investigation. The consistent of the trend and the plateau temperature in the caloric curves obtained with different thermometers suggests that the temperature is insensitive to the extraction methods.

\section*{VI. Summary}

The nuclear liquid-gas phase transitions under a constant volume and a constant pressure condition are investigated in the framework of the static AMD. Using the deuteron quadrupole momentum fluctuation thermometer, the caloric curves of fragmenting systems of $^{36}$Ar and $^{100}$Sn are extracted. A plateau structure of the caloric curves is obtained under a constant volume for those with density $\rho \leq$ 0.03 fm$^{-3}$ for both the two systems. A clear backbending in the caloric curve is observed under a constant pressure condition for all the pressures studied here. Distinct differences are observed for the signature of the liquid-gas phase transition measures under a constant volume or a constant pressure. The Similar behavior of caloric curves for $^{36}$Ar and $^{100}$Sn indicates there is no strong system size effect in the static AMD simulations.

The studies of light particle multiplicities indicate the $\alpha$ cluster enhancements at low excitation energy for all $r_{wall}$ investigated in the static AMD calculations. Signatures of the liquid-gas phase transition are observed in dM/dT and NVZ, but the signature is more prominent under a constant pressure and their temperature changes according to the density or pressure. The temperature extracted by the NVZ deviates from that of dM/dT and caloric curves under the constant volume condition, but the results under the constant pressure condition agree with those of dM/dT and caloric curves.

The experimental caloric curves are also compared with those of $^{100}$Sn of the static AMD simulation under a constant volume and a constant pressure conditions. The comparisons indicate that the liquid-gas phase transition occurs between a constant volume and a constant pressure, but more experimental studies are necessary to make a conclusive remarks. This study suggests that different measures for the liquid-gas phase transitions should be examined besides the caloric curves in order to draw a conclusion. The system size effect on the caloric curves in the experiments may come from the dynamic process during the heavy ion collisions, which still needs further investigation.

\section*{Acknowledgments}

The authors thank A. Ono for providing his code. This work is supported by the National MCF Energy R\&D Program of China (MOST 2018YFE0310200), the National Natural Science Foundation of China (Grant No. 11705242, 11805138 and 11905120) and the Fundamental Research Funds For the Central Universities (No. YJ201820, YJ201954) in China. This work is also supported by the US Department of Energy under Grant No. DE--FG02--93ER40773 and the Robert A. Welch Foundation under Grant A330.

\end{document}